# Electrical property tuning via defect engineering of single layer MoS$_2$ by oxygen plasma

Muhammad R. Islam,[a,b] Narae Kang [a,b], Udai Bhanu [a,b], Hari P. Paudel [a,b], Mikhail Erementchouk [a,b], Laurene Tetard [a,b], Michael N. Leuenberger [a,b,d], and Saiful I. Khondaker[a,b,c] *

[a] Nanoscience Technology Center, [b] Department of Physics, [c]School of Electrical Engineering and Computer Science, [d]College of Optics and Photonis (CREOL), University of Central Florida, Orlando, Florida 32826, USA.

* To whom correspondence should be addressed. E-mail: saiful@ucf.edu

We demonstrate that the electrical property of a single layer molybdenum disulfide (MoS$_2$) can be significantly tuned from semiconducting to insulating regime via controlled exposure to oxygen plasma. The mobility, on-current and resistance of single layer MoS$_2$ devices were varied up to four orders of magnitude by controlling the plasma exposure time. Raman spectroscopy, X-ray photoelectron spectroscopy and density functional theory studies suggest that the significant variation of electronic properties is caused by the creation of insulating MoO$_3$-rich disordered domains in the MoS$_2$ sheet upon oxygen plasma exposure, leading to an exponential variation of resistance and mobility as a function of plasma exposure time. The resistance variation calculated using an effective medium model is in excellent agreement with the measurements. The simple approach described here can be used for the fabrication of tunable two dimensional nanodevices on MoS$_2$ and other transition metal dichalcogenides.

## Introduction

The discovery of graphene by mechanical exfoliation from a layered bulk solid to a single atomic layer, and its extraordinary mechanical, electrical and optical properties have stimulated significant interests in other two dimensional (2D) materials and their heterostructures.[1-6] A unique aspect of the 2D materials is that they can exist with various intrinsic electronic properties such as metal, semimetal, semiconductor or insulator, which is defined by their co-ordination chemistry.[7] Among the different 2D materials, graphene (semimetal) and molybdenum disulfide (semiconductor) have been receiving the most attention. While graphene has a very high mobility, the lack of bandgap in its electronic structure limits its application in nanoelectronics and optical devices. On the other hand, MoS$_2$, a member of the transition metal dichalcogenides (TMDC) family, offers a layer dependent electronic bandgap: bulk MoS$_2$ is an indirect bandgap material with a gap of 1.2 eV which transitions into a direct bandgap of 1.8 eV in a single layer MoS$_2$.[8] As a result, field effect transistors (FETs) based on single layer MoS$_2$ have shown a current switching of up to $10^8$ [ref 9]. In addition, it was shown that the mobility of the MoS$_2$ FETs is also layer dependent: it increases with number of layers up to 10 layers and then decreases for higher thickness.[10] Consequently, single and multi-layer MoS$_2$ have been extensively studied and hold great promise for future nano-electronic devices. In fact, several prototype devices based on single and multilayer MoS$_2$ such as phototransistors, chemical sensors, photovoltaic devices, diodes, memory devices and integrated circuits have already been demonstrated.[9-19]

Electrical and optical property variations have been observed with number of layers, however, a more recent effort has been focused in engineering the material properties in the single layer MoS$_2$ by external control. The grand challenge is to determine whether the metal, semiconductor, and insulator 'phases' can be realized in the same 2D material using an external control without resorting to multiple 2D materials?  Since the 2D materials are atomically thin, their electronic and optical properties are highly sensitive to external control. For example, metallic behavior has been found in lithium intercalated single layer MoS$_2$.[20] Theoretical study predicts that, owing to its lower symmetry, strain engineering can be used to tune the MoS$_2$ bandgap, which can in turn modulate the electrical and optical properties. For example, it was predicted that a transition from direct to indirect bandgap can be achieved in single layer MoS$_2$ by strain engineering.[21-23]  Recent experiments suggest that electronic structure modification is indeed possible by strain engineering,[22] however experimental evidence of electrical property tunablility with such a technique is still lacking. The ability to continuously modulate the electronic properties in the same layer will enable the fabrication of nano-devices with a wide range of tunability of electronic and optical properties.

Here, we demonstrate a novel approach to continuously tune the electrical properties of single layer MoS$_2$ FET from semiconductor to insulator using an external control. The approach presented here is based on the controlled exposure of MoS$_2$ to oxygen plasma (O$_2$:Ar mixture of 20:80) for different exposure time, while keeping the single layer intact. We show that the mobility, on-current and resistance of a single layer MoS$_2$ FET vary exponentially by up to four orders of magnitude with respect to the plasma exposure time. Raman studies conducted before and after plasma treatment show a significant decrease of intensity of MoS$_2$ peaks with the creation of new MoO$_3$ peaks, while X-ray photoelectron spectroscopy (XPS) study also show peaks associated with MoO$_3$ after plasma exposure. We suggest that during exposure to oxygen plasma, the energetic oxygen molecules interact with MoS$_2$ and create MoO$_3$ rich defect regions, which are insulating. The coverage of the defect regions increases with exposure time. This picture is confirmed by density functional theory (DFT) calculations. We also used a theoretical model based on effective medium approximation to describe the resistance as a function of plasma exposure time. Our results show that the effective medium semiconductor (EMSC) made of MoS$_2$ including MoO$_3$ defect regions acts as a tunnel barrier for the injected conduction electrons, giving rise to the exponential increase in resistivity as a function of plasma exposure time, which is in excellent agreement with our experimental data. Our calculations demonstrate an increase in tunnel barrier height of 0.06 eV for each second of plasma exposure time. Our findings suggest a simple and efficient approach for the in-plane engineering of electrical properties of MoS$_2$ which can be applied to other TMDCs as well and will enable the fabrication of wide range of tunable 2D nanodevices.

**Experimental section**

*Device fabrication:* The devices were fabricated using single layer that were mechanically exfoliated from a commercially available crystal of molybdenite (SPI Supplies Brand, Natural Molybdenite) using adhesive tape micromechanical cleavage technique and deposited on a highly doped Si substrate capped with a thermally grown 250 nm thick SiO$_2$, Before MoS$_2$ deposition, the Si/SiO$_2$ wafers were cleaned using oxygen plasma followed by rinsing in acetone and isopropyl alcohol. Standard electron beam lithography (EBL) was used to pattern metal contacts on the MoS$_2$ flakes. First, a double layer electron beam resists, methyl methacrylate/poly(methyl methacrylate) (MMA/PMMA), was spun on the substrate and baked, followed by e-beam exposure and development in (1:3) methyl isobutyl ketone : isopropyl alcohol (MIBK:IPA). After defining the electrodes, 35 nm Au were deposited by thermal evaporation, followed by lift off in acetone.

*Electrical transport measurements:* The electron transport measurements of the $MoS_2$ device were performed in a probe station at ambient condition using a Keithley 2400 source meter and a DL instruments 1211 current preamplifier interfaced with LabView program. The measurements were performed before and after each oxygen plasma treatment. The plasma treatment on the $MoS_2$ devices was carried out using a commercial (Plasma Etch, PE-50) plasma chamber at a power of 100 W operating at 50 kHz. During plasma exposure, the pressure within the plasma chamber was held at 250 – 350 mTorr and a gas mixture of Oxygen (20%) and Argon (80%) flow at a constant rate of 15 sccm. For the first exposure, the samples were exposed for 2 s and subsequently they were exposed at 1 s interval and the electron transport measurements were repeated.

*Characterization:* XPS was performed to analyze any possible change in the chemical composition of $MoS_2$ falkes due to plasma treatment. XPS was carried out on Physical Electronics 5400 ESCA system utilizing a monochromatized Al Kα X-ray source. $MoS_2$ flakes containing both single layer and mulit-layer were exfoliated on $SiO_2$ substrate and XPS was carried out both before and after plasma exposure. The Raman spectra of the as exfoliated and plasma treated $MoS_2$ flakes were recorded with Witec alpha300RA confocal Raman system. The $MoS_2$ flakes were illuminated with 532 nm laser light in ambient air environment at room temperature. The power of the laser lines was kept below 1 mW in order to avoid any damage to the flake and on the other hand sufficient to obtain good signal to noise ratio.

*Theoretical calculation:* Density functional theory calculations (DFT) were performed to investigate the stability of MoOS and $MoO_3$ defects inside a single layer of $MoS_2$. For that we performed DFT calculations for a single layer of $MoS_2$, a single layer of $MoO_3$, a single layer of $MoS_2$ with MoOS defects, and a single layer of $MoS_2$ with $MoO_3$ defects. In each of the cases we considered a mesh of 9x9x1 k-points in the Brillouin zone. The ion-electron interaction is described by the projected augmented wave (PAW) method and the exchange-correlation energy is calculated using the Perdew, Burke and Ernzerhof (PBE) approximation within the framework of the generalized gradient approximation (GGA). The grid point cutoff of 415 eV is used and a maximum force of 0.1 eV/ Å on each atom is reached during the optimization process in all cases.

**Results and discussion**

Figure 1(a) shows an optical micrograph and Figure 1(b) show an atomic force microscopy (AFM) topography image of a representative single layer flake on $Si/SiO_2$ substrate. The flake's height profile shown in Figure 1(b) indicates a thickness of 0.9 nm, corresponding to a single layer.[17, 24] Number of $MoS_2$ layer was further confirmed by Raman spectroscopy taken before making electrical contacts to the flake, as shown in Figure 1 (c). Two prominent peaks at $E^2_g$ and $A^1_g$ corresponding to in plane and out of plane vibrations of Mo and S atoms were separated by a Raman position difference $\Delta = 19.28$ cm$^{-1}$ confirming the single layer nature of the flake.[25] Standard electron beam lithography (EBL) was used to pattern Au contacts on the $MoS_2$ flakes. Optical micrograph of a representative fabricated device is shown in the inset of Figure 1 (e).

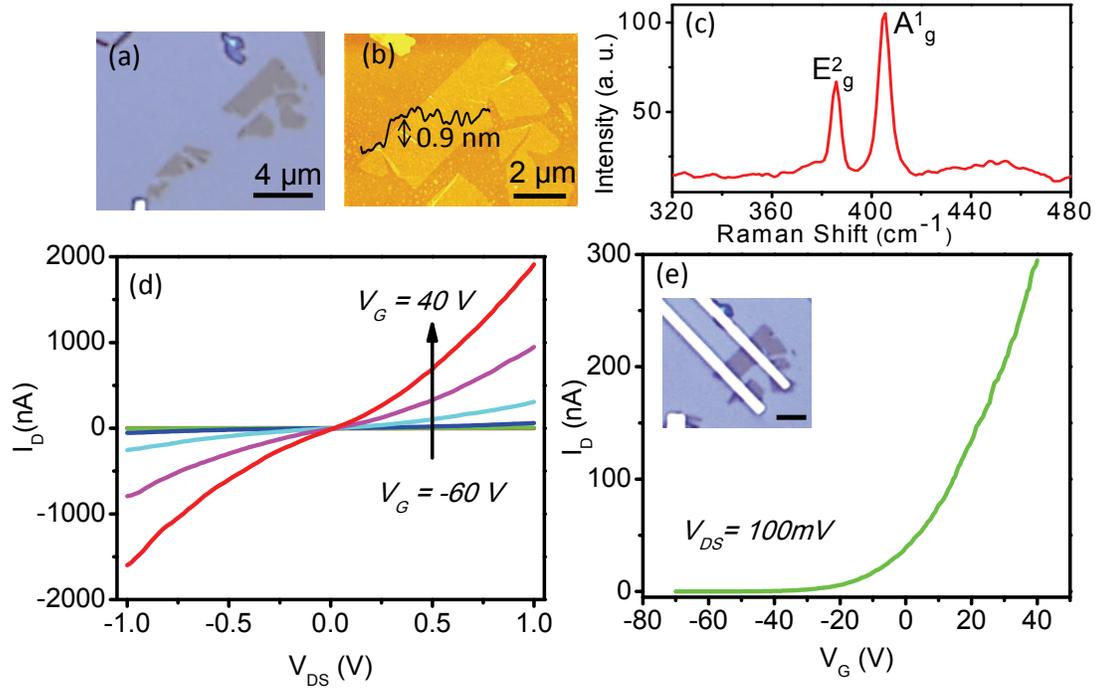

Figure 1. (a) Optical micrograph of a single layer MoS$_2$ flake exfoliated on Si/SiO$_2$ substrate. (b) AFM image (c) Raman spectrum of the single layer MoS$_2$. (d) Output characteristics of the single layer MoS$_2$ device at different back gate voltages ($V_G$) ranging from -60V to 40V with an interval of 20 V. (e) Transfer characteristics of the same device. The inset shows the optical micrograph of the device (scale bar is 2 μm).

Figure 1(d) shows the output characteristics ($I_D$ vs $V_{DS}$) for different back-gate voltages ($V_G$) varying from -60 to 40 V (bottom to top) with a step of 10 V. Increase of drain current with gate voltage indicates n-type FET behavior. Figure 1(e) shows the transfer characteristics (drain-current $I_D$ as a function of back-gate voltage $V_G$) measured at a fixed source-drain bias voltage $V_{DS}$ = 100 mV for the pristine MoS$_2$ device. The $I_D$ increased by several orders of magnitude with the increase of $V_G$. The current on-off ratio of the device is found to be ~$10^4$. The field effect mobility of the device was calculated to be 6 cm$^2$/Vs using the relation $\mu = (L/WC_GV_{DS})(dI_D/dV_G)$, where $L$ is the channel length, $W$ is the channel width and $C_G = \varepsilon_0\varepsilon_r A/d$ is the capacitance between the gate and SiO$_2$, with $\varepsilon_r$ ~$3.9\varepsilon_0$ is the effective dielectric constant of SiO$_2$, and $d$ (=250 nm) is the oxide thickness.[9]

Figure 2(a) shows the transfer characteristic of the same device after different plasma exposure time. For a unified view of the curves, we multiplied the curve for 0 sec (pristine MoS$_2$) exposure by 0.01 and 2 s exposure by 0.5. Interestingly, the drain current at all gate voltages decreases with increasing oxygen plasma exposure. This can be more clearly seen in Figure 2(b) (right axis) where the on-current at $V_G$ = 40 V is displayed in a semi-log scale. The drain current was ~285 nA for the as fabricated sample, which decreased exponentially with time to a value of less than 20 pA, a drop of more than four orders of magnitude, after only a total of 6 s plasma exposure time. After a 6 s exposure, the current becomes negligibly small. The mobility of the device after each plasma exposure is calculated from the $I_D$-$V_G$ curves in Figure 2(a), and is plotted in figure 2(b) (left axis) in a semi-log scale. Like the on-current, the mobility also drops exponentially from 6 $cm^2/Vs$ for the as fabricated sample to $4 \times 10^{-4}$ $cm^2/Vs$, after the 6 s plasma exposure. Similar to on-current, the decrease of mobility is also more than four orders magnitude with plasma exposure.

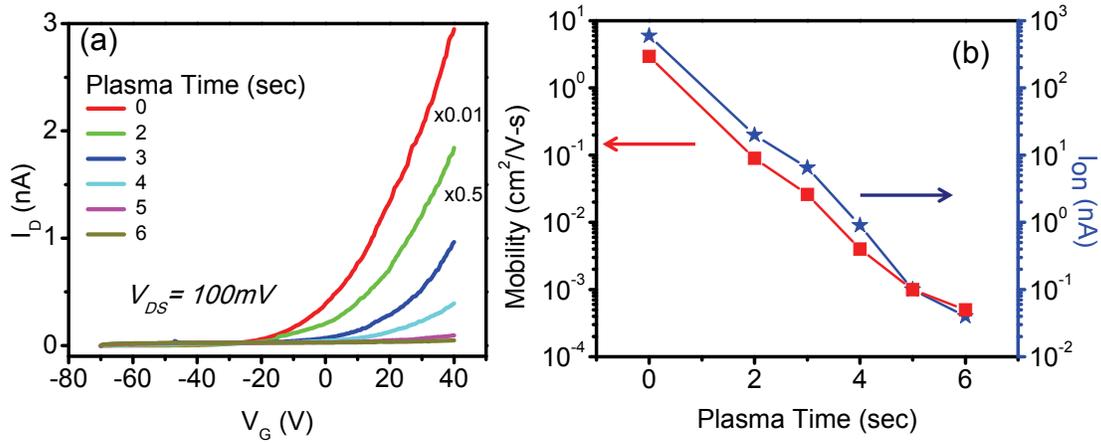

Figure 2. (a) Gate dependence of the source drain current ($I_D$) after different plasma exposure time. The curve corresponds to plasma exposure time of 0, 2, 3, 4, 5, and 6 sec respectively. (b) Effect of plasma exposure on the on-current (at $V_G = 40V$) and mobility of the single layer $MoS_2$ device.

Figure 3(a) shows the $I_D$-$V_{DS}$ graph of the device at $V_G = 40$ V for different plasma exposure time. It is observed that at all exposure times, the $I_D$-$V_{DS}$ curves are linear around the zero bias representing Ohmic behavior. Figure 3(b) demonstrates the dependence of resistance with respect to plasma exposure time. The resistance increased up to five orders of magnitude with increasing plasma exposure. The logarithmic plot in Figure 3 (b) demonstrates that the resistance increases exponentially upon plasma exposure. Similar changes in resistance were also observed for other gate voltages (see supporting information S1). This can be described by an effective medium model that shows that the exponential increase in the resistance as a function of plasma exposure time leads to the gradual increase of the tunnel barrier raised by the effective medium semiconductor (EMSC) material made of $MoS_2$ and strain-inducing $MoO_3$ rich defect regions.

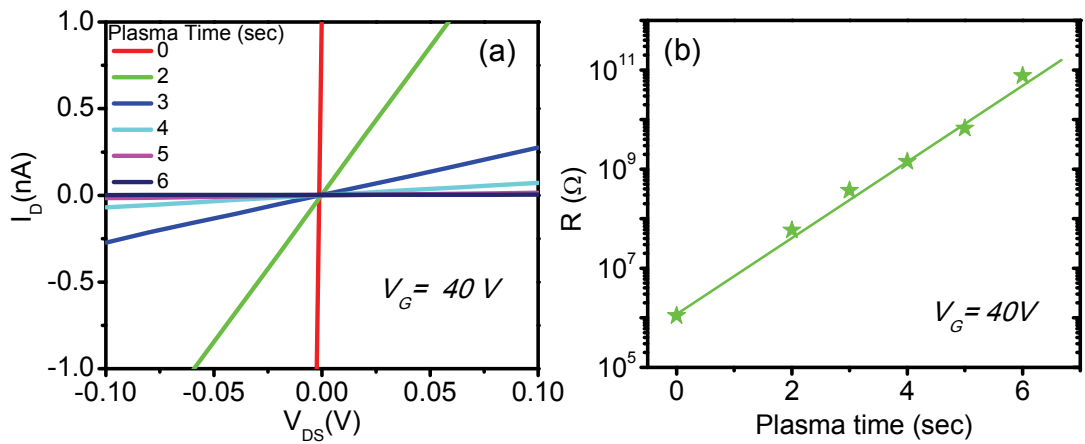

Figure 3. (a) $I_D$ vs $V_{DS}$ characteristics curve for the single layer $MoS_2$ device after different plasma exposure time. (b) Resistance of the device as a function of plasma exposure time. The green line is the linear fit of the logarithmic resistance as a function of exposure duration.

Similar device characteristics were obtained on two other single layer devices (see supporting information S2). To explore the physical mechanism responsible for the observed change in electronic transport properties, we performed Raman spectroscopy, and X-ray photospectroscopy (XPS) of the pristine and plasma treated MoS$_2$ flakes. Raman spectroscopy is a powerful tool to investigate changes in composition of 2D materials. Here we compare the Raman signature of the pristine flake and the plasma treated monolayer. Figure 4 (a) shows the Raman spectra of a representative single layer MoS$_2$ flake before (red curve) and after 6s of oxygen plasma treatment (green curve). The two Raman peak corresponding to $E_{2g}^1$ (~385 cm$^{-1}$) and $A_g^1$ (~410 cm$^{-1}$) modes, characteristic of MoS$_2$ observed in the pristine flake, clearly decrease in amplitude after treatment. Interestingly, $E_{2g}^1$ (in plane) is severely affected as a result of the treatment, while $A_g^1$ shifts only of 3 cm$^{-1}$ with a strong amplitude decrease (6 times) and a significant broadening. Finally the disappearance of the LAM mode at 450 cm$^{-1}$ also confirms the disruption of the MoS$_2$ lattice during oxygen plasma treatment. On the other hand, the appearance of other peaks observed in the 150-400 cm$^{-1}$ range indicates the formation of Mo-O bonds in the system, in particular at 225 cm$^{-1}$ corresponding to the B$_{3g}$ mode.[26]

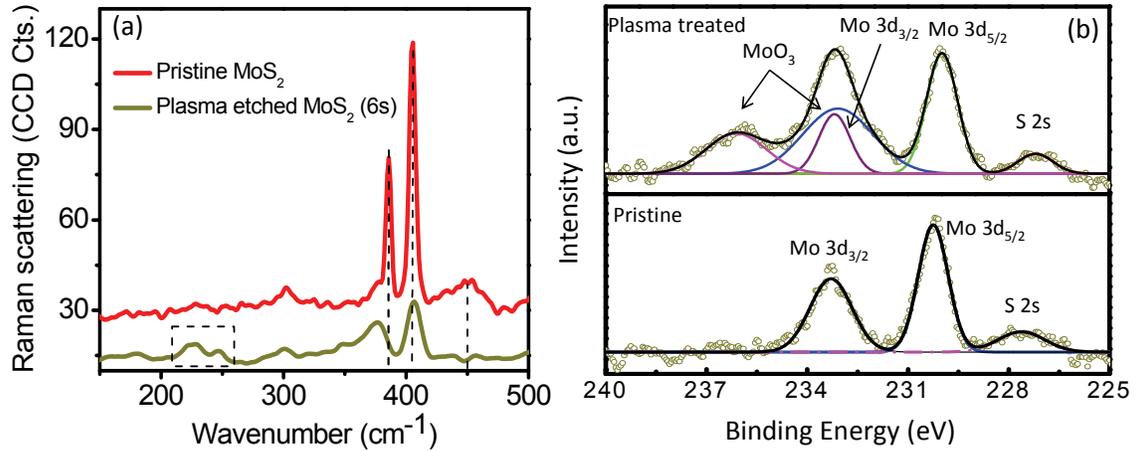

Figure 4. (a) Raman spectra of pristine MoS$_2$ (red) and plasma etched MoS$_2$ (green) obtained with a 532 nm excitation wavelength. The x-axis represents relative shift in wavenumber (rel. cm$^{-1}$). While MoS$_2$ modes were conserved (a), new Raman peaks corresponding to Mo-O bonds in MoO$_3$ could be measured in the flake after exposure. (b) X-ray photoelectron spectroscopy (XPS) of Mo (3d) and S (2s) core levels for pristine (lower panel) and plasma treated (upper panel) MoS2 flakes.

Figure 4(b) show the XPS spectra of pristine MoS$_2$ and plasma treated MoS$_2$ respectively. Three prominent peaks were observed at energies 227 eV, 229.7 eV, and 233.1 eV in pristine MoS$_2$ sample, origin of which has been attributed to binding energy of S 2s, Mo 3d$_{5/2}$ and Mo 3d$_{3/2}$ electrons in Mo-S bond of the MoS$_2$ crystal respectively.[27, 28] All these peaks were also found at same binding energies for the plasma treated sample, however an additional peak at energy 236.4 eV could be observed, corresponding to the higher oxidation state Mo$^{+6}$. This new peak further confirms the presence of MoO$_3$ in the plasma treated sample.[28] In previous experiments, oxygen plasma was used on single layer and multi-layered MoS2 for etching purpose.[18, 29] In order to ensure that the electrical property variation is not due to etching, we have acquired SEM image after completing the measurements and found that the single layer remained intact after 6 s plasma treatment (see supplementary information S3). The difference may be due to the process recipe. During the plasma process, the etching rate depends on the oxygen concentration, rf power, frequency and plasma exposure time.[30, 31] The previous experiments used high power reactive ion etcher while we used a $O_2$:Ar mixture of 20:80, moderately low power and low frequency plasma for short duration.

Based on the Raman, and XPS studies, we propose the following qualitative picture to explain the electrical property evolution of $MoS_2$: during plasma treatment high energetic charge particles bombard on the $MoS_2$ surface. Since S atoms have a smaller mass compared to Mo, S atoms can move out of the lattice site and lattice vacancies are created. Because of the excess oxygen supplied by the plasma, oxidation takes place at the defect sites created by S vacancies on the surface.[32, 33] The oxidation process can be described as $2MoS_2 + 7O_2 \rightarrow 2MoO_3 + 4SO_2$.[34] $MoO_3$ has an experimentally measured bandgap of 3.2-3.8 eV, making it insulating in nature.[35] Therefore, the creation of $MoO_3$ in $MoS_2$ creates significant distortion of lattice (Figure 5) which increases with increasing plasma exposure time. The resulting material can be treated as an EMSC which can explain the exponential variation of resistance (discussed later).

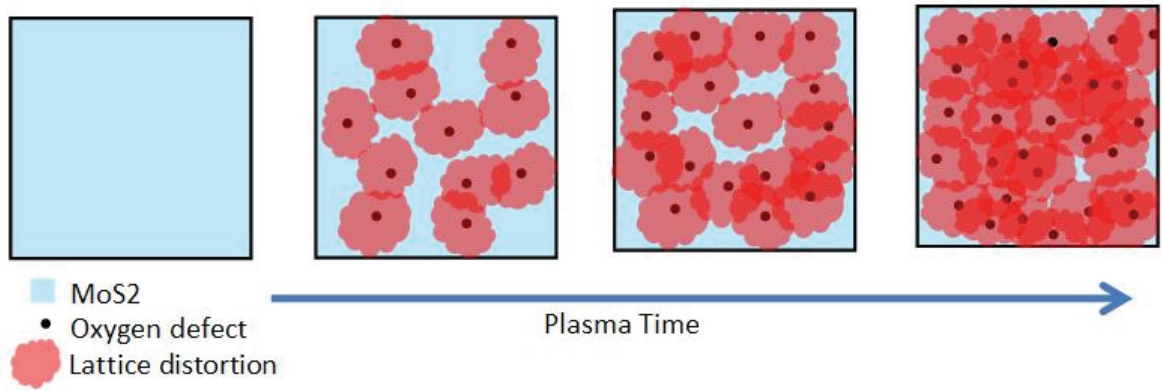

Figure 5. Structural model of electrical property tuning via defect engineering in $MoS_2$ single layer as a function of oxygen plasma exposure time. With the increase of plasma exposure time, insulating $MoO_3$ domains are formed in $MoS_2$. The $MoO_3$ rich defect regions not only change the atomic sites locally, where they replace the S atoms, but also lead to lattice distortions. With increasing plasma exposure, the distortion also increases.

In the following we present our calculations based on density functional theory (DFT) as implemented in the atomistix tool kit (ATK) program (see materials and methods). The most important fact that needs to be remembered is that a single layer of $MoS_2$ and a single layer of $MoO_3$ have completely different lattice structures. A 2D layer of $MoS_2$ has Mo atoms sandwiched by the S atoms with honeycomb lattice structure as shown in the Figure 6(a). Two (0001) $MoS_2$ layers bind weakly through van der Waals interaction, making mechanical exfoliation along the c direction possible. On the other hand, $MoO_3$ has orthorhombic layered structure with separate layers stacked along the b direction. A single sheet of $MoO_3$ consists of bilayers with both sides terminated with O atoms, as shown in the Figure 6(b). The interlayer is bridged by the O atoms along the [100] direction. Due to the O atom termination of the single sheet of $MoO_3$ on both sides, there is a weak van der Waals bond that connects two sheets, as in the case of $MoS_2$, but along the [010] direction. Thus, the introduction of $MoO_3$ defects in a single layer of $MoS_2$ results in significant lattice distortions. Our results show that $MoO_3$ rich defect regions can be created when oxygen is introduced and that $MoO_3$ defects are stable inside $MoS_2$ (Figure 6), which is in line with the identification of $MoO_3$ in the Raman spectra in Figure 4 (a). Figures 6 (c), (d), (e), and (f) show the change in the lattice structures in the $MoS_2$ 2D sheet when few oxygen O (red) atoms replace the covalently bonded S (yellow) atoms. First we replace one S atom by an O atom in the 2D unit cell (Figure 6c), and optimize the structure. We find that the original plane of $MoS_2$ is distorted and the O atom is shifted by a distance of 0.32 Å along the c direction after optimization (Figure 6(d)). In addition, the O atom covalently bonds with the Mo (grey) atom by forming a molybdenum oxysulphide (MoOS). We then replace three S atoms by three O atoms within a 2x2 supercell (doubling the lattice constant of the unit cell along the a and b directions). Figure 6 (e) and (f) show the configuration before and after optimization, respectively. As O atoms form bonds with the Mo atom, the original plane of the S atoms is distorted. After optimization, it is evident that the covalent bonds form between one of the Mo atoms and the three O atoms, providing evidence for the stable configuration containing a $MoO_3$ defect. This result clearly shows that when O atoms interact with $MoS_2$, $MoO_3$ defect regions can form and create structural distortions of the lattice in all directions. The formation of $MoO_3$ is in agreement with our Raman and XPS data while structural deformation leads to an increase of resistance and decrease of mobility. This will be further discussed in a later section.

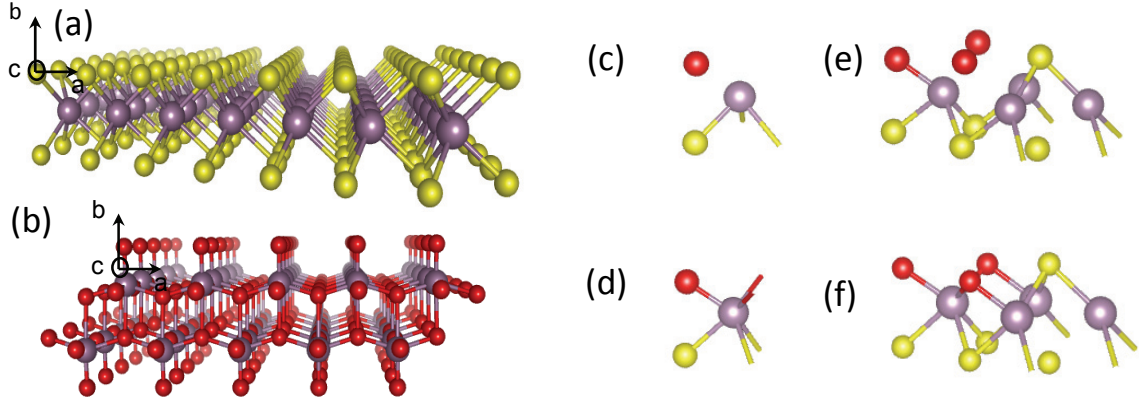

Figure 6. A single sheet of (a) $MoS_2$ and (b) $MoO_3$. Such sheets are weakly attached by the van der Waal bond along [001] direction in $MoS_2$ and [010] direction in $MoO_3$. (c) A single O atom (red) is replaced by a single S atom (yellow) in the unit cell (d) Unit cell structure after optimization. (e) Three O atoms replace three S atoms in a 2x2 supercell. (f) After optimization it is evident that the covalent bonds form between one of the Molybdenum atoms (grey) and the three O atoms, providing evidence for the stable configuration containing a $MoO_3$ defect.

The DFT analysis presented in Figure 6 shows conclusive evidence about the presence of $MoO_3$ defects along with their surrounding lattice distortions due to oxygen plasma. It is clear that the effect of the plasma treatment on the structure of the $MoS_2$ layer is not homogeneous, it forms regions with variable concentration of $MoO_3$ defect regions in $MoS_2$ and forms a complex network of heterojunctions that affect the electron transport (Figure 6). For a qualitative description of the effect of such complex network on resistance, we adopt the approximation of an effective medium, which regards the plasma-treated material as an EMSC (see supplementary figure S4). For simplicity, we assume the EMSC has a homogeneous structure whose work function depends on plasma exposure time ($\tau$). Thus we effectively have a heterostructure of EMSC - MoS$_2$. The electron transport through EMSC - MoS$_2$ interface is determined by the respective band bending and the built-in potential $\Delta\phi(\tau)$, which depends on the concentration of defects and, therefore, depends on the plasma exposure time. It is natural to assume that with an increasing concentration of defects the built-in potential evolves towards its limiting value characterizing MoS$_2$ - MoO$_3$ interface ($\tau \rightarrow \infty$). Thus the maximal value of $\Delta\phi$ can be estimated using Anderson's rule:

$$\Delta\phi_{max} = \phi_{MoO_3} - \phi_{MoS_2} \approx 2 \text{ eV}, \tag{1}$$

where $\phi_{MoS_2} \approx 4.9$ eV [36] and $\phi_{MoO_3} \approx 6.9$ eV [37] are respective work functions. The values of $\phi_{MoS_2}$ and $\phi_{MoO_3}$ reported in the literature vary noticeably but in any case $\Delta\phi_{max} > 1$ eV can be expected. Such strong built-in potential results in great increase of resistance $R_{max}/R(0) \sim \exp(\Delta\phi_{max}/k_B T) \sim 10^{12}$, where $k_B$ is the Boltzmann constant and $T$ is temperature (see Supplemental materials for details). This is well above the values of ~$10^4$ measured in the experiment and suggest that our final EMSC is not crystalline MoO$_3$ sheet in agreement with physical characterization. Rather, our experiment suggests that $\Delta\phi$ varies with the plasma exposure time and should have values: $0 \leq \Delta\phi(\tau) \ll \Delta\phi_{max}$. Taking this circumstance into account we can expand $\Delta\phi(\tau)$ in series with respect to $\tau$ and keeping only the linear term we present $\Delta\phi(\tau) = \alpha\tau$, where $\alpha$ is the rate at which the barrier increases. Thus we obtain

$$\ln(R(\tau)/R(0)) \sim \frac{\alpha\tau}{k_B T} \tag{2}$$

for the dependence of resistance on the plasma exposure time. It is interesting to note that the result obtained from this rather simple model is in excellent agreement with our experimental results shown in Figure 3 (b), with $\alpha \approx 6 \cdot 10^{-2}$ eV·s$^{-1}$. This calculation also suggest that with longer oxygen plasma exposure, the work function of the exposed region increases. For the same reason, the mobility and on-current decreases with plasma exposure time.

**Conclusions**

In conclusion, we report the electrical property tunability of single layer MoS$_2$ devices upon controlled exposure to oxygen plasma. The on-current, mobility and resistance can be tuned up to four orders of magnitude by varying the plasma exposure time. Based on our Raman spectroscopy, XPS, and DFT studies we present strong evidence that the significant decrease of mobility and on-conductance is caused by the creation of insulating MoO$_3$-rich disordered domains which causes significant lattice distortion in the MoS$_2$ sheet upon oxygen plasma exposure. , Using a simple effective medium model we show that the tunnel barrier of the plasma treated MoS$_2$ increases at a rate of $\alpha \approx 6 \cdot 10^{-2}$ eV·s$^{-1}$ with plasma exposure time. The method of electrical property tuning of MoS$_2$ devices described here can serve as an enabling technology for fabricating tunable 2D nanodevices for electronic and optoelectronic applications.

*Acknowledgment.* M.N.L. acknowledges support from NSF (grant ECCS-0901784), AFOSR (grant FA9550-09-1-0450), and NSF (grant ECCS-1128597).

*Supporting Information Available:* (1) Resistance versus plasma time for Vg = 20 V. (2) Effect of oxygen plasma on a second single layer device. (3) Scanning electron microscope image of plasma exposed MoS$_2$ flake. (4) Theoretical calculation of resistance. This material is available free of charge via internet at http://pubs.rsc.org.

**References**

1. A. K.Geim, I. V. Grigorieva, *Nature* 2013, **499**, 419-425.
2. M. S. Xu, T. Liang, M. M. Shi and H. Z. Chen, *Chemical Reviews* 2013,**113**, 3766-3798.
3. Q. H. Wang, K. Kalantar-Zadeh, A. Kis, J. N. Coleman and M. S. Strano, *Nature Nanotechnology* **2012**, 7, (11), 699-712.
4. S. Z. Butler, S. M. Hollen, L. Cao, Y. Cui, J. Gupta, H. R. Gutierrez, T. F. Heinz, S. S. Hong, J. Huang, A. F. Ismach, E. Johnston-Halperin, M. Kuno, V. V. Plashnitsa, R. D. Robinson, R. S.; Ruoff, S. Salahuddin, J. Shan, L. Shi, M. G. Spencer, M. Terrones, W. Windl, and J. E. Goldberger, *ACS Nano* 2013, **7**, 2898-2926.
5. V. Nicolosi, M. Chhowalla, M. G. Kanatzidis, M. S. Strano, and J. N. Coleman, *Science* 2013**, 340**, 1420.
6. D.Jariwala, V. K. Sangwan, L. J. Lauhon, T. J. Marks, and M. C. Hersam, *ACS Nano* 2014, **8**, 1102-1120.
7. M.Chhowalla, H. S. Shin, G.Eda, L.-J.Li, K. P. Loh and H. Zhang, *Nature Chemistry* 2013, **5**, 263-275.
8. K. F. Mak, C. Lee, J. Hone, J. Shan and T. F. Heinz, *Physical Review Letters* 2010, **105**, 136805.
9. B. Radisavljevic, A. Radenovic, J. Brivio, V. Giacometti and A. Kis, *Nature Nanotechnology* 2011, **6**, 147-150.
10. S. Das, H. Y. Chen, A. V. Penumatcha and J. Appenzeller, *Nano Letters* 2013, **13**, 100-105.
11. Z. Y. Yin, H. Li, L. Jiang, Y. M. Shi, Y. H. Sun, G. Lu, Q. Zhang, X. D. Chen and H. Zhang, *ACS Nano* 2011, **6**, 74-80.
12. W. Zhang, C. Chuu, J. Huang, C. Chen, M. Tsai, Y. Chang, C. Liang, Y. Chen, Y. Chueh, J. He, M. Chou and L. Li, *Scientific Reports* 2014, **4**, 3826.
13. F. K. Perkins, A. L. Friedman, E. P. M. Cobas, Campbell, G. G. Jernigan, and B. T. Jonker, *Nano Letters* 2013, **13**, 668-673.
14. D. J. Late, Y. K. Huang, B. Liu, J.Acharya, S. N. Shirodkar, J. J. Luo, A. M. Yan, D. Charles, U. V. Waghmare, V. P. Dravid and C. N. R. Rao,. *ACS Nano* 2013, **7**, 4879-4891.
15. M. Fontana, T. Deppe, A. K. Boyd, M. Rinzan, A. Y. Liu, M. Paranjape and P. Barbara, *Scientific Reports* 2013, **3**, 1634.
16. B. Radisavljevic, M. B. Whitwick and A. Kis, *ACS Nano* 2011, **5**, 9934-9938.
17. H. Wang, L. L. Yu, Y. H. Lee, Y. M. Shi, A. Hsu, M. L. Chin, L. J. Li, M. Dubey,J. Kong, and T. Palacios, *Nano Letters* 2012, **12**, 4674-4680.
18. M. K. Chen, H. Nam, S. J. Wi, L. Ji, X. Ren, L. F. Bian, S. L. Lu and X. G. Liang, *Applied Physics Letters* 2013**, 103**, 142110.
19. M. K. Chen, H. Nam, S. Wi, G. Priessnitz, I. M. Gunawan, and X. G. Liang, *ACS Nano*, (Article ASAP).
20. G. Eda, T. Fujita, H. Yamaguchi, D. Voiry, M. Chen and M. Chhowalla, *ACS Nano* 2012, **6**, 7311-7317.
21. P. Johari and V. B. Shenoy, *ACS Nano* 2012, **6**, 5449-5456.

22. K. He, C. Poole, K. F. Mak, and J. Shan, *Nano Letters* 2013, **13**, 2931-2936.
23. A. Castellanos-Gomez, R. Roldan, E. Cappelluti, M. Buscema, F. Guinea, H. S. J. van der Zant, and G. A. Steele, *Nano Letters* 2013, **13**, 5361-5366.
24. M. Buscema, M. Barkelid, V. Zwiller, H. S. J. van der Zant, G. A. Steele and A. Castellanos-Gomez, *Nano Letters* 2013, **13**, 358-363.
25. C. Lee, H. Yan, L. E. Brus, T. F. Heinz, J. Hone and S. Ryu, *ACS Nano* 2010, **4**, 2695-2700.
26. M. A. Py, P. E Schmid, and J. T. Vallin, *Nuovo Cimento Della Societa Italiana Di Fisica B-General Physics Relativity Astronomy and Mathematical Physics and Methods* 1977, **38**, 271-279.
27. P. A. Spevack AND N. S. McIntyre, *Journal of Physical Chemistry* 1993, **97**, 11031-11036.
28. J. Yang, S. Kim, W.Choi, S. H. Park, Y. Jung, M. H. Cho and H. Kim, *ACS Applied Materials & Interfaces* 2012, **5**, 4739-4744.
29. B. Radisavljevic, and A. Kis, *Nature Materials* **2013**, 12, (9), 815-820.
30. S. Manolache, M.Sarfaty and F. Denes, *Plasma Sources Science & Technology* 2000, **9**, 37-44.
31. X. K. Lu,H. Huang, N. Nemchuk and R. S. Ruoff, *Applied Physics Letters* 1999, **75**, 193-195.
32. N. M. D. Brown, N. Y. Cui and A. McKinley, *Applied Surface Science* 1998, **134**, 11-21.
33. J. R. Lince and P. P. Frantz, *Tribology Letters* 2000, **9**, 211-218.
34. B. C. Windom, W. G. Sawyer, and D. W. Hahn, *Tribology Letters* 2011, **42**, 301-310.
35. Y. J. Lee, W. T. Nichols, D.-G. Kim and Y. Do Kim, *Journal of Physics D-Applied Physics* 2009, **42**, 115419.
36. R. Schlaf, O. Lang, C. Pettenkofer and W. Jaegermann, *Journal of Applied Physics* 1999, **85**, 2732-2753.
37. J. Meyer, S. Hamwi, M. Kroeger, W. Kowalsky, T. Riedl and A. Kahn, *Advanced Materials* 2012, **24**, 5408-5427.

**Supporting Information**

# Electrical property tuning via defect engineering of single layer MoS$_2$ by oxygen plasma

Muhammad R. Islam,[a,b] Narae Kang [a,b], Udai Bhanu [a,b], Mikhail Erementchouk [a,b], Laurene Tetard [a,b], Michael N. Leuenberger [a,b,d], and Saiful I. Khondaker[a,b,c] *

[a] Nanoscience Technology Center, [b] Department of Physics, [c] School of Electrical Engineering and Computer Science, [d] College of Optics and Photonis (CREOL), University of Central Florida, Orlando, Florida 32826, USA.

* To whom correspondence should be addressed. E-mail: saiful@ucf.edu

## 1. Resistance versus plasma time for Vg = 20 V:

Figure S1(a) shows the $I_D$-$V_{DS}$ graph of the device at $V_G$ = 20V for different plasma exposure

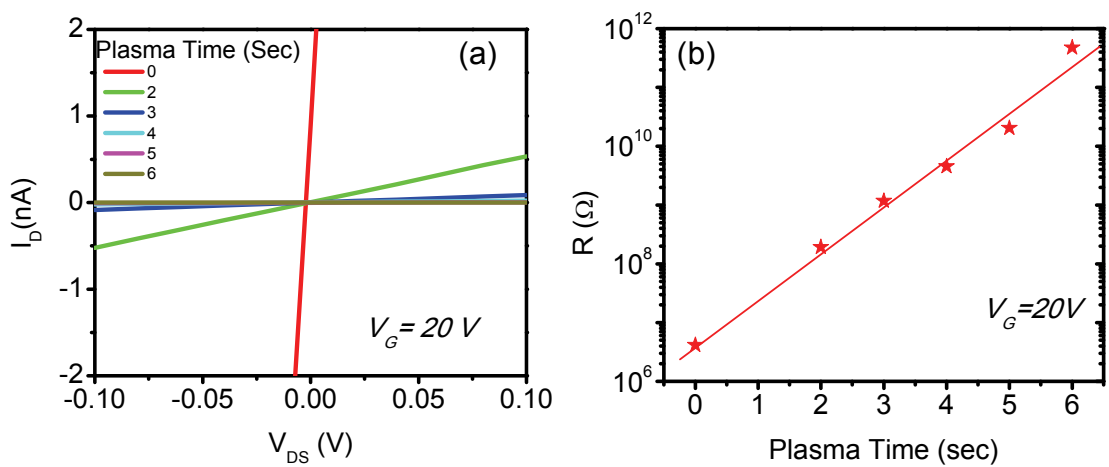

**S1.** (a) $I_D$ vs $V_{DS}$ characteristics curve for the single layers MoS$_2$ device at different plasma exposure at $V_G$=20V. (b) Resistance of the device as a function of plasma exposure time at $V_G$=20V time. Figure S1(b) shows the dependence of resistance on the plasma exposure time. A large variation in resistance is observed after plasma exposure which increased exponentially with the plasma exposure time.

## 2. Effect of oxygen plasma on a second single layer device:

Transfer characteristic of the a second single layer device after different plasma exposure are shown in Figure S2(a). For the convenience of comparison and to obtain a unified view of the curves, we multiplied the curve for 0 sec exposure by 0.01. The on current at $V_G$=40 V is displayed in a semi-log scale in Figure S2(b) (right axis). The drain current varies exponentially with time from ~76 nA for the as fabricated sample to value of less than 15 pA for 6 sec plasma exposure. Similar to the on-current, the mobility also drops exponentially from 7.3 cm$^2$/Vs for as fabricated sample to 1.6x10$^{-4}$ cm$^2$/Vs, after the 6 s plasma exposure. Figure S3(c) shows the dependence of with plasma exposure time. A five order increase in resistance is observed with

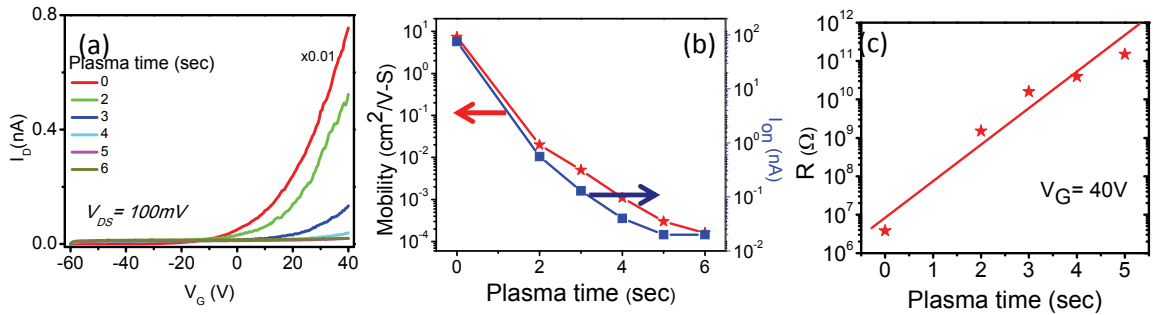

**S2.** (a) Gate dependence of the source drain current ($I_D$) for another single layer device. The curve corresponds to plasma exposure time of 0, 2, 3, 4, 5, 6, sec respectively. (b) Effect of plasma exposure on the on current (at $V_G$=40V) and mobility of the single layers MoS$_2$ device. (c) Resistance of the device as a function of plasma exposure time.

the plasma exposure time.

### 3. Scanning electron microscope image of plasma exposed MoS$_2$ flake:

To check any possible change in the surface morphology of the plasma exposed MoS$_2$ flake we have taken the Scanning electron microscope (SEM) image of flake after plasma exposure. Figure S3 shows the SEM image of a plasma exposed single layer MoS$_2$ flake. No signature of etching was observed from the figure.

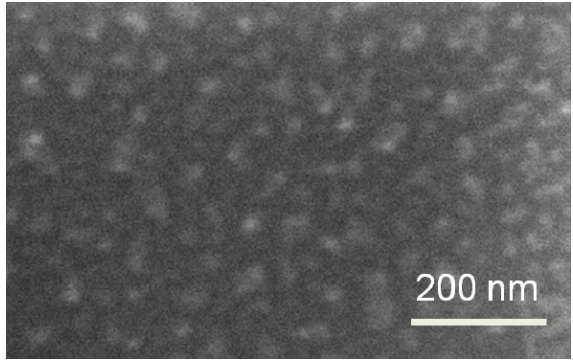

**S3.** SEM image of plasma exposed single layer MoS$_2$ flake.

### 4. Theoretical calculation of resistance:

We considered the plasma-treated material as an effective-medium semiconductor (EMSC). This is shown in figure S4, where right side figure show that the intact MoS$_2$ underneath the gold electrode formed a heterojunction with EMSC. The current through a heterojunction with a relatively high built-in potential $\Delta\phi$ has a form similar to the Shockley diode equation,

$$I = I_S \left(e^{eV/nk_BT} - 1\right) \approx I_S \frac{eV}{nk_BT} \quad \text{------- (3)}$$

where $V$ is applied voltage, $k_B$ is the Boltzmann constant, $T$ is temperature, $n$ is the ideality factor. The dependence on the band mismatch at the heterojunction enters this equation through

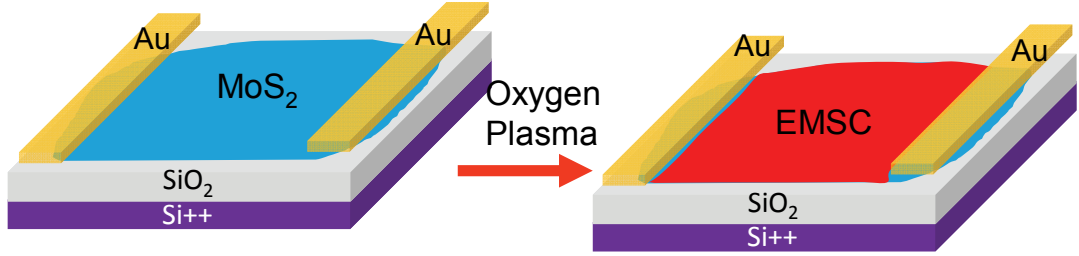

**S4.** Schematic of creating of EMSC region upon plasma exposure showing MoS2 – EMSC heterojunction.

the saturation current

$$I_S(\Delta\phi) \approx A T^2 \exp\left(-\frac{\Delta\phi}{k_B T}\right),$$

where $A = em^* k_B^2/(2\pi^2 \hbar^3)$ is the Richardson constant. In order to define an effective resistance of the heterojunction we consider the limit of small applied voltage, $V \ll V_T = nk_B T/e \approx 0.2$ V (for $T = 400$ K and $n = 1$) and rewrite Eq. (3) as $I = V/R(\Delta\phi)$, where $R(\Delta\phi) = V_T/I_S(\Delta\phi)$. Thus we find

$$\ln[R(\Delta\phi)] = C + \frac{\Delta\phi}{k_B T}.$$

This formula provides the dependence of total resistance on $\Delta\phi$ in the case when the junction under consideration makes the major contribution.